\documentstyle[fleqn]{jsc}

\title{Computer Algebra in Spacetime Embedding }

\author{WALDIR L. ROQUE\thanks{E-Mail: ROQ@LNCC.BITNET} \\
                \mbox{} \\
       {\em Departamento de Matem\'atica,} \\
       {\em Universidade de Bras\'\i lia,} \\
       {\em 70910 -- Bras\'\i lia, DF, Brasil} \\
                \mbox{} \\
                      AND \\
                \mbox{} \\
       RENATO P. DOS SANTOS\thanks{E-Mail: RPS@LNCC.BITNET} \\
                \mbox{} \\
       {\em Centro Brasileiro de Pesquisas F\'\i sicas,} \\
       {\em Rua Xavier Sigaud, 150,} \\
       {\em 22290 -- Rio de Janeiro, RJ, Brasil}}

\begin{document}

\maketitle
   
\begin{abstract}
In this paper we describe an algorithm  to determine
the vectors  normal  to a space-time V$_{4}$  embedded in a
pseudo-Euclidean manifold M$_{4+n}$. An application of this algorithm is
given considering the Schwarzchild spacetime geometry embedded
in a 6 dimensional pseudo-Euclidean mani\-fold,  using the algebraic
computing system  REDUCE.
\end{abstract}

\section{Introduction}

The General Relativity Theory defines the physical spacetime as
a differentiable pseudo-Riemmanian 4-dimensional manifold
(Hawking \& Ellis, 1973) . The spacetime is seen through the
{\em intrinsic}  geometry. However, many interesting  results
have shown that the {\em extrinsic} geometry gives sometimes a
better understanding of the physical structure of the
spacetime. As an example of its  growing interest we can cite the recent 
studies of Gravity Theories in more than four dimensions
and the study of the geometry of extended objects as in String
and Membrane Theories. 

	In contrast to General Relativity, where the metric uniquely specifies
the geometry, to describe a spacetime locally and isometrically embedded in a
pseudo-Euclidean manifold with dimension $4 + n$, two new
quantities have to be considered: the {\em second fundamental
form} and the {\em torsion} vector (or the third
fundamental form). These two quantities are
well known in the study of differentiable manifolds (Eisenhart,
1946) and come out from the Gauss-Codazzi-Ricci equations,
which are the integrability conditions for the existence of the
embedding (Maia, 1986).
 
	Rigorously it is not necessary to know the embedding to obtain the second
fundamental form and the torsion vector . Both can be obtained directly through the
field equations (Maia \& Roque, 1989) which are highly
non-linear parcial differential equations in the 4 spacetime
coordinates. However, if we know the embedding {\em a priori}, to
determine these two quantities we need to get first the set of
vectors normal to the spacetime V{$_{4}$}.

	In this paper we will be concerned with the discussion  of an
algorithm which helps to determine these vectors from the
embedding coordinates and then to find out the second fundamental
form and the torsion vector. The following section sets up the
main equations that rule the embedding theory. In section 3 the
algorithm for determining the vectors normal to the spacetime
V{$_{4}$} is described.  This algorithm has been
implemented in the algebraic computing system REDUCE and its
application is done in section 4 for the case of Schwarzchild
embedding. Some comments and remarks on an extension of this
algorithm to select and determine the rank of any $m \times n$
matrix are left to section 5 of the paper.

\section{Embedding Equations }  

	A local embedding of a spacetime V{$_{4}$} in a
pseudo-Euclidean M{$_{4+n}$} manifold is done when a set of
Cartesian coordinates  $X^{\mu}$ is specified as functions of
the spacetime coordinates $x^i$ ( greek indices run from 1 to
$4+n$, lowercase latin letters run from 1 to 4 and capital latin
letters run from 5 to $4+n$). At any point of the manifold we
can find a set of n vector fields  $N_A$ orthogonal to
V{$_{4}$} and to themselves. Thus if $\eta{_{\mu\nu}}$ denotes
the Cartesian components of the metric of M{$_{4+n}$}, then the
following set of equations are valid
\vskip 5pt
\hbox to \hsize{$X{^\mu_{,i}} X{^\nu_{,j}} \eta_{\mu\nu} = g{_{ij}}$; \hfill (1.a)}  
\vskip 5pt
\hbox to \hsize{$X{^\mu_{,i}} N{_A^\nu} \eta_{\mu\nu} = 0$ ; \hfill (1.b)} 
\vskip 5pt
\hbox to \hsize{$N{_A^\mu} N{_B^\nu} \eta_{\mu\nu} = g_{AB}$; \hfill (1.c)}
\vskip 5pt
\addtocounter{equation}{1}
where $g_{ij}$ is the spacetime metric, 
$g_{AB} = k^{2} \epsilon^{A} \delta_{AB}$, $\epsilon^{A} = \pm 1$,         
depending on the signature of  $M_{4+n}$,
$X{^\mu_{,i}}={\partial X^\mu \over \partial x^i}$ are the
components of the tangent vectors to V{$_{4}$} (partial derivatives 
with respect to spacetime coordinates are indicated by a comma, as usual) 
and $k$ is a constant.

The second fundamental form and the torsion vector are given, respectively, by
\vskip 5pt
\hbox to \hsize{$b_{ijA} =  N^{\mu}_{A} X^{\nu}_{,ij}\eta_{\mu\nu}, 
\quad  b_{ijA} = b_{jiA},$ \quad  {\rm and} \hfill(2.a)}
\vskip 5pt
\hbox to \hsize{$A_{iAB} = N^{\mu}_{A,i} N^{\nu}_{B}\eta_{\mu\nu},\quad A_{iAB} = 
-A_{iBA}$. \hfill(2.b)}
\vskip 5pt
\addtocounter{equation}{1}

	In a matricial form the set of equations (1.b) can be written as
\begin{equation}
{\bf S}\cdot {\bf Y}{_A} = 0\, ,\qquad A=5,...,4+n;
\end{equation}
where {\bf S} is the $4\times (4+n)$ matrix formed by the components
of the tangent vectors to V{$_{4}$} multiplied by the metric components
of M{$_{4+n}$} and  {\bf Y}${_A}$ is the column matrix $(4+n)\times
1$ formed by the  components of the vectors $N_A$.

	The homogeneous system described by equation (3) can be 
solved (for the non-trivial solution) by
taking into account pure algebraic considerations: we need to
find a square submatrix of {\bf S} of order $4\times 4$ that is
invertible. That is always possible as the rows of the matrix
{\bf S} are exactly the components of the vectors that generates
the  tangent space of the spacetime. Thus, they are linearly
independent. Therefore from linear algebra we know that there
exist a submatrix $ 4\times 4$ of {\bf S} that is non-singular.

	Let {\bf P} be a $4\times 4$ submatrix of {\bf S} that is invertible
and {\bf Q} the matrix formed from {\bf S} taking out the
elements of {\bf P}. {\bf Q} is a $4\times n$ matrix. The system (3)
can be written in the equivalent form
\begin{equation}
{\bf P}\cdot {\bf{\bar Y}}{_A} + {\bf Q}\cdot {\bf {\bar {\bar Y}}}{_A} = 0,
\end{equation}
where ${\bf {\bar Y}}{_A}$  are the components of
${\bf Y}_A$ associated to the invertible submatrix  and  ${\bf
{\bar{\bar Y}}}{_A}$ the components of ${\bf Y}{_A}$ associated
to the  remaining columns. Thus, from (4) we have that
\begin{equation}
{\bf P}\cdot{\bf {\bar Y}}{_A}  = - {\bf Q}\cdot {\bf {\bar{\bar Y}}} {_A},
\end{equation}
which allows us to write,
\begin{equation}
{\bf {\bar Y}}{_A} = - {\bf P}^{-1}\cdot {\bf
Q}\cdot {\bf {\bar{\bar Y}}} {_A}.
\end{equation}

	Taking into account the above definitions we  write in the following
section an algorithm to determine these quantities explicitly.

\section{The Algorithm}  

\begin{center}{\scriptsize ALGORITHM A}\end{center}

\begin{program}
   \item[A1:] Given the set of $4+n$ Cartesian coordinates $X^\mu$ as a
$(4+n)\times 1$ column matrix and the metric tensor $\eta_{\mu\nu}$
as a $(4+n)\times (4+n)$ square matrix, compute the {\bf S} matrix as
${\bf S}_{i\nu}=X{^\mu_{,i}}\eta_{\mu\nu}$. 
   \item[A2:] Using the algorithm B, decompose {\bf S} and ${\bf Y}{_A}$ matrices
in submatrices {\bf P}, {\bf Q}, ${\bf {\bar Y}}{_A}$ and ${\bf
{\bar {\bar Y}}}{_A}$ such that ${\rm det}({\bf P}) \not= 0$ and
such that
${\bf {\bar Y}}{_A}$ contains the components of $N_A$
corresponding to {\bf P} and ${\bf {\bar {\bar Y}}}{_A}$ those
corresponding to {\bf Q}.
   \item[A3:] Substitute {\bf P}, {\bf Q}, ${\bf {\bar Y}}{_A}$ 
and ${\bf {\bar {\bar Y}}}{_A}$ in 
${\bf {\bar Y}}{_A}-{\bf P}{^{-1}}\cdot {\bf Q}\cdot
{\bf {\bar {\bar Y}}}{_A}  = 0$,
obtaining a system of four linear equations in those components of $N_A$
corresponding to {\bf P}.
   \item[A4:] Solve that system of equations for the four components of each $N_A$ 
in ${\bf {\bar Y}}{_A}$ in terms of the $n$ others.
   \item[A5:] {\bf for} $A \leftarrow 5,\ldots,4+n$ {\bf do} 
              {\bf for} $B \leftarrow 5, \ldots, A$ {\bf do}
   \item[ ]\begin{program}
       \item[A5a:] Substitute the expressions  for $N_A$ in
${\bf Y}{_A}\eta {\bf  Y}{_B}  = g_{AB}$
obtaining a non-linear equation in the components of $N_A$ corresponding 
to {\bf Q}.
       \item[A5b:] Solve this equation for one of the remaining components of 
 $N_A$ in terms of the others.
       \item[A5c:] Return this solution to the next equation generated in step A5a. 
At the end of loop  $n(n-1)/2$ components of $N_A$ will remaing
arbitrary. {\bf endfor}
   \end{program}
   \item[A6:] Compute the second fundamental form and the torsion vector from
eqns. (2.a) and (2.b). {\bf stop}
\end{program}

\begin{center}{\scriptsize ALGORITHM B}\end{center}

\begin{program}
   \item[B1:] (Inicialization) $p_1\leftarrow 1$ ($p_1$ points to a candidate to 
         be the first column of {\bf P} (or ${\bf {\bar Y}}$)).
   \item[B2:] {\bf while} $p_1\le n+1$ {\bf and} ${\rm det}({\bf P})=0$\footnote{
Note that ${\rm det}({\bf P})$ can result an
expression that can be zero or not depending of physical informations
unavaliable to REDUCE.  In this case, if it is not immediatly zero, the
actual program could ask the user if it should be taken as zero by use
of the internal (symbolic) procedure YESP.}  {\bf do}
   \item[ ] {\bf begin} $p_2\leftarrow p_1+1$ ($p_2$ points to a candidate to be the second 
         column of {\bf P} (or ${\bf {\bar Y}}$).
   \item[ ] {\bf while} $p_2\le n+2$ {\bf and} ${\rm det}({\bf P})=0$ {\bf do}
   \item[ ] {\bf begin} $p_3\leftarrow p_2+1$ ($p_3$ points to a candidate to be the third 
         column of {\bf P} (or ${\bf {\bar Y}}$).
   \item[ ] {\bf while} $p_3\le n+3$ {\bf and} ${\rm det}({\bf P})=0$ {\bf do}
   \item[ ] {\bf begin} $p_4\leftarrow p_3+1$ ($p_4$ points to a candidate to be the 
         fourth column of {\bf P} (or ${\bf {\bar Y}}$).
   \item[ ] {\bf while} $p_4\le n+4$ {\bf and} ${\rm det}({\bf P})=0$ {\bf do}
   \item[ ] {\bf begin} $j\leftarrow 1$, $k\leftarrow 1$ ($j$ points to a column 
         of {\bf S} (or {\bf Y}), $k$ to a column of {\bf Q} 
         (or ${\bf {\bar {\bar Y}}}$)).  
   \item[ ]\begin{program}
        \item[ ] {\bf repeat} 
            \item[ ] {\bf begin} {\bf if} $j=p_1$ {\bf then} store the column of {\bf S} pointed by
             $j$ as the first column of ${\bf P}$, the one of {\bf Y} 
             as the first of ${\bf {\bar Y}}$ {\bf else}  
        \item[ ] {\bf if} $j=p_2$ {\bf then} store the column of {\bf S} pointed by 
             $j$ as the second column of ${\bf P}$, the one of {\bf Y} 
             as the second of ${\bf {\bar Y}}$ {\bf else}  
        \item[ ] {\bf if} $j=p_3$ {\bf then} store the column of {\bf S} pointed by 
             $j$ as the third column of ${\bf P}$, the one of {\bf Y} 
             as the third of ${\bf {\bar Y}}$ {\bf else}  
        \item[ ] {\bf if} $j=p_4$ {\bf then} store the column of {\bf S} pointed by 
             $j$ as the fourth column of ${\bf P}$, the one of {\bf Y} 
             as the fourth of ${\bf {\bar Y}}$ {\bf else}

        \item[ ] Store the column of {\bf S} pointed by $j$ as the $k$-th column 
             of ${\bf Q}$, the one of {\bf Y} as the $k$-th column of 
             ${\bf {\bar {\bar Y}}}$ and $k\leftarrow k+1$ {\bf endif}.
        \item[ ] $j\leftarrow j+1$ {\bf end}
        \item[ ] {\bf until} $j>n+4$ {\bf endrepeat}. \end{program} 
   \item[ ] $p_4\leftarrow p_4+1$ {\bf endwhile}
   \item[ ] $p_3\leftarrow p_3+1$ {\bf endwhile}
   \item[ ] $p_2\leftarrow p_2+1$ {\bf endwhile}
   \item[ ] $p_1\leftarrow p_1+1$ {\bf endwhile}. 
   \item[B3:] {\bf return} ${\bf P}$, ${\bf Q}$, ${\bf {\bar Y}}$, and 
${\bf {\bar {\bar Y}}}$. 
\end{program}
 
	The termination of the algorithm at step B3 is guaranteed by
the existence of the  non-singular submatrix {\bf P}.

	The algorithm above has been implemented in the algebraic
computing system REDUCE (Hearn, 1986; Rayna, 1987; Stauffer et. al.,
1988) making use of its  MATRIX facilities (see Davenport et.
al., (1988), for a good introduction to  matrix representation
in Computer Algebra). However, for shortage of space, we left the program
out of the paper.

\section{The Schwarzchild Embedding}

	In the specific case of Schwarzchild spacetime the
embedding (Rosen, 1965) is
done in a 6-dimensional pseudo-Euclidean manifold ($n=2$) with
metric $\eta_{\mu\nu} = diag(-1,-1,+1,+1,+1,+1)$. 
The Schwarzchild embedding is given by the coordinates,
\begin{eqnarray*}
X^1 & = & \sqrt\beta\, {\cos t}, \\
X^2 & = & \sqrt\beta\, {\sin t}, \\
X^3 & = & f(r), \\
X^4 & = & r \sin \theta \cos \phi, \\
X^5 & = & r \sin \theta \sin \phi, \\
X^6 & = & r \cos \theta,
\end{eqnarray*}
where $\beta=\beta(r) \quad (\beta(r)=1-{2m\over r})$,
$f(r)$ is a well defined function of $r$, and $r$, $\theta$, $\phi$, and $t$
denote the spacetime coordinates. The four vectors tangent to
V{$_{4}$} are determined taking the derivative of the
coordinates $X^{\mu}$ with respect to each one of the spacetime
coordinates. We denote by $N{_A}{^\mu}$, $\mu=1,...,6$ the
components of the normal vectors with $A=5,6$, respectively. To
determine these vectors the following conditions have to be
considered: 
i) the orthogonality of the normal vectors with
respect to the tangent vectors (eq. 1.b) (from this we obtain a
set of 8 equations) and 
ii) orthonormality of the normal vectors (eq. 1.c)(from this we 
get 3 equations).

	Out of a total of 11 equations we have now to determine the 6
components of the two vectors $N_5$ e $N_6$. We have a set
of 11 equations for 12 unknowns. Notice that our unknowns are
functions of the spacetime coordinates.

	According to the algorithm (and program) developed in the
previous section, we just need to set $n=2$ and ask REDUCE
to calculate the matrix {\bf S} (step A1). After some algebraic
manipulation we obtain for the Schwarzchild spacetime embedding
the normal vectors,
\begin{eqnarray*}
N{_5^\mu} & = & h(r) ( {\cos t \over \sqrt{\beta}}, 
{\sin t \over \sqrt{\beta}}, -\beta, 0, 0, 0) \\
N{_6^\mu} & = & l(r) ( -{\cos t \over \sqrt{\beta}},
 -{\sin t \over \sqrt{\beta}}, {4mf' \over \beta^2},
 - ({4mf'^2 \over \beta^2}-{\beta \over 4m}) \cos\phi
\sin\theta, \\
 & & - ({4mf'^2 \over \beta^2}-{\beta \over 4m}) \sin\phi
\sin\theta, - ({4mf'^2 \over \beta^2}-{\beta \over 4m}) \cos\theta) \\
\end{eqnarray*}
where $ h(r) = {4mf'\sqrt{\beta} \over \sqrt{\beta^3-16m^2f'^2}}$, \quad  
$l(r) = {4m\beta^2 \over \sqrt{16m^2f'^2-\beta^3}\sqrt{16m^2(1+f'^2)-\beta^3}}$ \quad 
and  $f' = {df\over dr}$.  

	It is easy now to  calculate  with REDUCE  (but tedious by hand)  the 
{\em second fundamental form } and the {\em torsion} vector, 
from the eqs.(2.a and 2.b - step A6):
\vskip5pt
\noindent $b_{115}=-h(r)$,\quad $b_{116}=-l(r)$,
\quad
 $b_{225}=-{\beta \over 4m}({f'' \over f'} + {3\beta \over 4m}) h(r)$,
\vskip3pt
\noindent $b_{226}=({4m^2 \over \beta^2}f'f'' + {3\beta^2 \over 16m}) l(r)$, 
\quad
 $b_{336}={r \over 4m\sqrt{\beta}f'}h(r)$,
\vskip3pt
\noindent $b_{446}= {r \over 4m\sqrt{\beta}f'}\sin^2 \theta h(r)$,
\vskip3pt
\noindent $A_{256}={4mf'' + 3\beta f' \over \sqrt{\beta(\beta^3-16m^2f'^2)}} l(r)$.

\section{Final Remarks}  

	The geometrical and physical analysis of these quantities are
not  the main concern of this paper. However it is important to point
out that
geometrically the second fundamental form and the torsion vector
are fundamental quantities as they determine, together with the
metric, the structure of the embedding manifold and
physically if General Relativity
has to be considered as part of a more general theory of embedded
manifolds then, besides the metric which represents the
classical gravitation, these two quantities have also to be 
considered: the second fundamental form  may be
interpreted as the source of the matter fields and the torsion vector
may represent a Yang-Mills  gauge field ( see Maia, 1986, for 
details).

	The calculations were initially done in interactive form  with
the version  3.2 of REDUCE 
 running in an IBM PC-XT and later on (by demand) in a
microVAX running VMS. Finally we coded a fairly general program\footnote{
Complete program and output listings may be obtained from the authors.}  
for the 3.3 version of REDUCE requiring only as input the number
of extra dimensions $n$, the set of Cartesian coordinates
$X^{\mu}$, and the metric $\eta_{\mu \nu}$ of the embedding
manifold  $M_{4+n}$.

	The available physical memory of the PC
(640 Kb, but less when the system is loaded) is a great
limiting factor for the  execution of calculus with more general
functions and/or higher dimensions (this would involve
matrices with order greater then $4\times 6$). To circumvent the very
often problem of free storage cell explosion in the PC, we had to
make the trick of using the output of the results as input for
the following steps.  Though this initial limitation at the PC,
the problem above would have been far more difficult to solve
with paper and pencil than with the interactive use of REDUCE.

	The algorithm developed here can be extended to determine all
non-singular submatrices of a given matrix determining, in
addition, its rank\footnote{If {\bf A} is a $m\times n$ matrix, the
number of submatrices of order $k\times k$ of {\bf A} is given by
$N={m\choose k} {n\choose k}$, where $k \le min(m,n)$ and
${a\choose b} = {a!\over (a-b)! b!}$.}. Thus it can also be used
to establish the existence and type of solution of a system of
linear equations  by the simple  analysis of its
coefficients' matrix  and  extended matrix ranks, for either
symbolic (functions) or numeric matrix entries, as the
manipulation is purely algebraic.

\section*{Acknowledgments}

W.L.Roque is grateful to the CNPq for financial support through a
research grant. 

\section*{References}

	Davenport, J. H., Siret, Y., Tournier, E. (1988). {\it
Computer Algebra: Systems and Algorithms for Algebraic
Computation}. Academic Press. 

	Eisenhart, L. P. (1946). {\em Riemannian Geometry}. Princeton
University Press.

	Hawking, S. W. \& Ellis, G. F. R. (1973).{\em The Large Scale
Structure of Space-Time}. Cambridge University Press.

	Hearn, A. C. (1986). {\em REDUCE Manual}. The Rand Corporation,
Santa Monica.

	Maia, M. D. (1986).	The Physics of the
Gauss-Codazzi-Ricci Equations. {\em Mat. Aplic.  Comp. {\bf 5},
283-292}; On Kaluza-Klein Relativity. {\em Gen. Rel. Grav. {\bf 18}, 695-699}.

	Maia, M. D. \& Roque, W. L. (1988). Classical Membrane Cosmology. 
{\em Phys. Lett. A {\bf 139}, 121-124}.

	Rayna, G. (1987). {\em REDUCE: A Software for Algebraic
Computation},   Springer-Verlag.

	Rosen, J. (1965). Embedding of Various Relativistic
Riemannian Spaces in Pseudo-Euclidean Spaces. {\em Rev. Mod. Phys.
{\bf 37}, 204-214}.

	Stauffer, D., Hehl, F.W., Winkelmann, V. and Zabolitzky, J.G.
(1988). {\em Computer Simulation and Computer Algebra}, Springer-Verlag.  

\end{document}